\documentclass[12pt]{iopart}

\usepackage[dvips]{graphicx}
\usepackage{subfig}

\def\bra#1{\mathinner{\langle{#1}|}} 
\def\ket#1{\mathinner{|{#1}\rangle}}

\begin{document}

\title[Non-Markovian dynamics of the spin-boson model]{A quantum jump description for the non-Markovian dynamics of the spin-boson model}

\author{E.-M. Laine}

\address{Department of Physics and Astronomy, University of Turku, FI-20014 Turku, Finland}
\ead{emelai@utu.fi}
\begin{abstract}
We derive a time-convolutionless master equation for the spin-boson model in the weak coupling limit. The temporarily negative decay rates in the master equation indicate short time memory effects in the dynamics which is explicitly revealed when the dynamics is studied using the non-Markovian jump description. The approach gives new insight into the memory effects influencing the spin dynamics and demonstrates, how for the spin-boson model the the co-operative action of different channels complicates the detection of memory effects in the dynamics.
\end{abstract}

\maketitle

\section{Introduction} \label{Introduction}
The spin-boson model describes a two-level system interacting with a bosonic environment. It is widely studied, mostly because of its many applications in the field of quantum information theory and chemistry. The model can describe for example a simple two-well potential \cite{Leggett}, double quantum dots \cite{Brandes} and different biomolecular systems \cite{Garg, Gilmore}. Because the spin-boson model is not exactly solvable many approximative approaches have been used in order to unravel the dynamics of the spin-boson system. The model has been studied using perturbation theory in the system-environment coupling with the Markov approximation \cite{Cheche, Wilhelm} and projection operator methods \cite{Loss}. In reference \cite{Leggett} an approximative scheme called the interacting blip approximation (NIBA) has been introduced. It is a perturbative approach in the tunneling matrix element based on path integral methods. There also exists an exact approach to the problem using Feynman-Vernon path integral method \cite{Grifoni1, Grifoni2}. However, this presents only a formal solution.

Understanding the short-time dynamics of the spin-boson model is crucial in understanding quantum computing systems \cite{Privman, Privman2}. The Markovian approximation is valid only for long times and thus the non-Markovian description of the dynamics is needed. NIBA fails for large tunneling matrix element and cannot therefore be applied to biomolecular systems \cite{Garg, Gilmore}. Many projection operator methods and path integral methods are difficult to treat and do not necessarily give much physical insight to the problem. The main motivation of this paper is to write down a time-convolutionless master equation describing the dynamics of the spin-boson model and to study the equation via non-Markovian quantum jumps. This description gives a clear picture of the processes involved in the dynamics and reveals the role of memory effects in the evolution of the spin.

In section \ref{TCL} we derive a time-convolutionless master equation describing the reduced dynamics of the spin-boson model by using the time-convolutionless projection operator method \cite{Breuer}. The master equation is derived in second order and the secular approximation is performed. The derived master equation is then studied in section \ref{NMQJ} via the non-Markovian quantum jump (NMQJ) approach \cite{Piilo, piilo2, piilo3} and the conclusions are presented in section \ref{conclusions}.

\section{Derivation of the master equation} \label{TCL}
The Hamiltonian describing the spin-boson model is 
\begin{equation}
	H=\frac{\epsilon}{2}\sigma_z+\frac{\Delta}{2}\sigma_x+\sum_n{\omega_n a_n^\dagger a_n}+\frac{\sigma_z}{2} \sum_n{\lambda_n(a_n^\dagger+a_n)}.
	\label{equation1}
	\end{equation} 
Here, $\sigma_z=\ket{e}\bra{e}-\ket{g}\bra{g}$, $\sigma_x=\ket{e}\bra{g}+\ket{g}\bra{e}$ in the eigenbasis of the two-state system $\left\{\ket{e},\ket{g}\right\}$, $\epsilon$ is the energy bias between the two states, $\Delta$ is tunneling amplitude between the states, and $\omega_n$ and $a_n$ are the energies and annihilation operators of the corresponding modes of the bosonic environment respectively. The tunneling amplitude $\Delta$ is assumed to be a real number. The parameter $\lambda_n$ describes the coupling strength between the two-state system and the $n$th mode of the reservoir. The coupling of the system to the environment is a bilinear coupling through operator $\sigma_z$ ,i.e., the environment is sensitive to the value of the operator $\sigma_z$.

We start the derivation of the master equation by writing down a general form of the second order time-convolutionless master equation \cite{Breuer}
\begin{equation}
\frac{\rmd}{\rmd t}\tilde{\rho}_\mathcal{S}(t)=-\int_0^t{\tr_\mathcal{E}[\tilde{H}_I(t),[\tilde{H}_I(t'),\tilde{\rho}_\mathcal{S}(t)\otimes \rho_\mathcal{E}]]\rmd t'}.
	\label{equation2}
\end{equation}
Here, the notation $\tilde{A}(t)$ stands for an operator $A$ written in the interaction picture. Now, if one writes a decomposition for the interaction Hamiltonian of the form $H_I=S\otimes E$ in terms of the system eigenoperators $S_\omega$, i.e. $\tilde{H}_I(t)=\sum_\omega{\rme^{-\rmi \omega t}S_\omega \otimes \tilde{E}(t)}$, and performs the secular approximation, \eref{equation2} may be written in the form
\begin{equation}
\dot{\tilde{\rho}}_\mathcal{S}(t)=-\rmi[H_{LS}(t),\tilde{\rho}_\mathcal{S}(t)]+\mathcal{D}(t)(\tilde{\rho}_\mathcal{S}(t)),
	\label{equation3}
\end{equation}
with
\begin{equation}
\mathcal{D}(t)(\tilde{\rho}_\mathcal{S}(t))=\sum_\omega{\gamma_\omega(t)\left[S_\omega\tilde{\rho}_\mathcal{S}(t)S^\dagger_\omega-\frac{1}{2}\left\{S^\dagger_\omega S_\omega,\tilde{\rho}_\mathcal{S}(t) \right\}\right]}
	\label{equation4}
\end{equation}
and
\begin{equation}
H_{LS}(t)=\sum_\omega{\lambda_\omega(t)S^\dagger_\omega S_\omega}.
	\label{equation5}
\end{equation}
The time-dependent coefficients $\gamma_\omega(t)$ and $\lambda_\omega(t)$ in \eref{equation4} and \eref{equation5} are determined from the expression 
\begin{eqnarray}
	\Gamma_\omega(t)&\equiv& \int_0^t{\rmd t' \rme^{\rmi\omega t'}\tr_\mathcal{E}\left\{\tilde{E}^\dagger(t)\tilde{E}(t-t')\rho_\mathcal{E}\right\}}\nonumber \\
	&=&\int_0^t{\rmd t' \rme^{\rmi\omega t'}\tr_\mathcal{E}\left\{\tilde{E}^\dagger(t')\tilde{E}(0)\rho_\mathcal{E}\right\}}
	\label{equation6}
	\end{eqnarray}
by the relation $\Gamma_{\omega}(t)=\frac{1}{2}\gamma_\omega(t)+\rmi \lambda_\omega(t)$.

The system eigenoperators for the spin-boson model written in the system eigenbasis $\left\{\psi_+,\psi_-\right\}$ are
\begin{equation}
	S_0=\frac{\epsilon}{2\omega_0}\sigma_z,\quad S_{\omega_0}=-\frac{\Delta}{2\omega_0}\sigma_-,\quad S_{-\omega_0}=-\frac{\Delta}{2\omega_0}\sigma_+,
	\label{equation7}
\end{equation}
where $\sigma_z=\ket{\psi_+}\bra{\psi_+}-\ket{\psi_-}\bra{\psi_-}$, $\sigma_-=\ket{\psi_-}\bra{\psi_+}$, $\sigma_+=\ket{\psi_+}\bra{\psi_-}$ and $\omega_0=\sqrt{\epsilon^2+\Delta^2}$ is the system eigenenergy. In the limit of continuous spectrum of the environment and for zero temperature one can write the coefficients $\gamma_\omega(t)$ in \eref{equation4} as
\begin{equation}
	\gamma_{\omega}(t)=2\int_0^t{\rmd t'\int_0^\infty{\rmd \omega'J(\omega')\cos\left[(\omega-\omega')t'\right]}},
	\label{equation8}
\end{equation}
where $J(\omega)$ is the spectral density. For an Ohmic spectral density $	J(\omega)=\frac{\alpha}{2}\omega \rme^{-\omega/\omega_c}$ \cite{Breuer}, where $\omega_c$ is the cutoff frequency, the coefficients are
\begin{eqnarray}
		\gamma_{\pm \omega_0}(t)&=&\frac{\alpha\omega_c}{1+\omega_c^2t^2}\left[\omega_c t\cos(\omega_0 t)\mp\sin(\omega_0 t) \right]\nonumber \\
		&+&\alpha\omega_0 \rme^{\mp\omega_0/\omega_c}\left[\Re\left[\mathrm{Si}(z)\right]\mp \Im\left[\mathrm{Ci}(z)\right]\pm \frac{\pi}{2} \right]\nonumber \\
	\gamma_0(t)&=& \frac{\alpha \omega_c^2 t}{1+\omega_c^2 t^2},
	\label{equation9}
\end{eqnarray}
where $\mathrm{Ci}(z)$ is the cosine integral, $\mathrm{Si}(z)$ is the sine integral and $z=\omega_0t+\rmi \frac{\omega_0}{\omega_c}$. The decay rates $\gamma_{\omega_0}(t)$ and $\gamma_{-\omega_0}(t)$ have oscillatory behavior and they get temporarily negative values resembling memory effects in the reduced dynamics, while the decay rate $\gamma_0(t)$ remains always positive. For simplicity we ignore the Lamb shift \eref{equation5} and restrict to the regime $\omega_c\ll \omega_0$, where the secular approximation is valid in the non-Markovian regime.

The solution to the master equation presented in \eref{equation3}, \eref{equation7}and \eref{equation9} can be written in terms of the density matrix elements as
\begin{equation}
	\rho(t)=M(t)\rho(0),
	\label{equation10}
\end{equation}
with
\begin{equation}
M(t)=\left(\begin{array}{cccc}
	g(t)&0&0&f(t)\\
	0&e^{-\zeta(t)}&0&0\\
	0&0&e^{-\zeta(t)}&0\\
	1-g(t)&0&0&1-f(t)
\end{array} \right).	\label{equation11}
\end{equation}
Here, 
\begin{eqnarray}
	g(t)&=&f(t)+e^{-\eta(t)},\nonumber\\
	f(t)&=&e^{-\eta(t)}\xi(t),\nonumber\\
	\eta(t)&=& \frac{\Delta^2}{4\omega_0^2}\int{(\gamma_{\omega_0}(t)+\gamma_{-\omega_0}(t))\rmd t},\nonumber\\
	\xi(t)&=&\frac{\Delta^2}{4\omega_0^2}\int{\gamma_{-\omega_0}(t)e^{-\eta(t)}\rmd t},\nonumber\\
\zeta(t)&=&\int{(\frac{\epsilon^2}{2\omega_0^2}\gamma_0(t)+\frac{\Delta^2}{8\omega_0^2}(\gamma_{\omega_0}(t)+\gamma_{-\omega_0}(t)))\rmd t}.
	\label{equation12}
\end{eqnarray}
The matrix in equation \eref{equation10} acts on $\rho(0)$ as on a column vector consisting of the density matrix elements.

\section{Jump description of the dynamics} \label{NMQJ}
The master equation derived in the previous section can be studied with the NMQJ-method \cite{Piilo}, which is a generalization of the Monte Carlo wave function method \cite{Molmer} to the non-Markovian regime. The key increment in this description is that whenever the coefficients $\gamma_\omega(t)$ get negative values the direction of the jump process is reversed for each wave function realization in the corresponding channel. The density matrix for an ensemble of wave functions with $N$ members at time $t$ is 
\begin{equation}
	\rho_\mathcal{S}(t)=\sum_{\alpha}\frac{N_\alpha(t)}{N}\ket{\phi_\alpha(t)}\bra{\phi_\alpha(t)},
	\label{equation13}
\end{equation}
where $N_\alpha(t)$ is the number of ensemble members in state $\ket{\phi_\alpha(t)}$ at time $t$. When the decay rates are positive the jumps taking place in a single wave function history can be determined by the Monte Carlo wave function method \cite{Molmer}. When a decay rate becomes negative, the reversed jump operator in the corresponding channel $j$ is $D_{\alpha\to\alpha'}^j(t)=\ket{\phi_{\alpha'}(t)}\bra{\phi_\alpha(t)}$, with $	\ket{\phi_\alpha(t)}=C_{j_-}\ket{\phi_\alpha'(t)}/||C_{j_-}\ket{\phi_\alpha'(t)}||$. The reversed jumps can take the wave functions to superposition states destroyed earlier by jumps and they therefore resemble the memory effects in the dynamics. The memory effects, by recreation of superpositions, allow a reversal of the decoherence process which we call here recoherence. The reversed transition occurs during a time step $\delta t$ with probability
\begin{equation}
	P_{\alpha\to\alpha'}^j(t)=\frac{N_{\alpha'}(t)}{N_{\alpha}(t)}|\gamma_j(t)|\delta t\bra{\phi_{\alpha'}(t)}C_j^\dagger C_j\ket{\phi_{\alpha'}(t)}.
	\label{equation14}
\end{equation}

Let us study the dynamics of the spin-boson model via the jump description. We concentrate here on giving a clear picture of the effect of the reversed jumps on the dynamics and will not perform the actual unraveling of the master equation \eref{equation3}. The intention is to give insight into the origin of memory effects by describing the system dynamics in terms of non-Markovian jumps. The plots describing the dynamics presented in this section are obtained by numerically evaluating \eref{equation11}. 

\begin{figure}[!h]
    \includegraphics[width=0.4\textwidth]{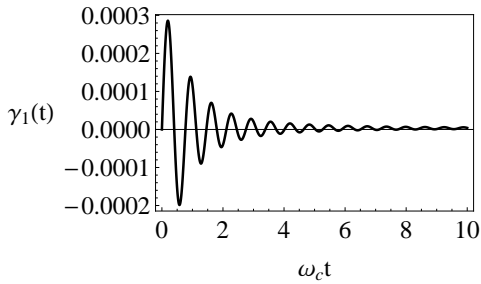}
  \caption{The decay rate $\gamma_1(t)$ in units of $\omega_c$ for $\alpha=0.01$, $\omega_0/\omega_c=10$ and $\epsilon/\Delta=1/(2\sqrt{3})$.}
  \label{figure1}
\end{figure}

\begin{figure}[!h]
    \includegraphics[width=0.4\textwidth]{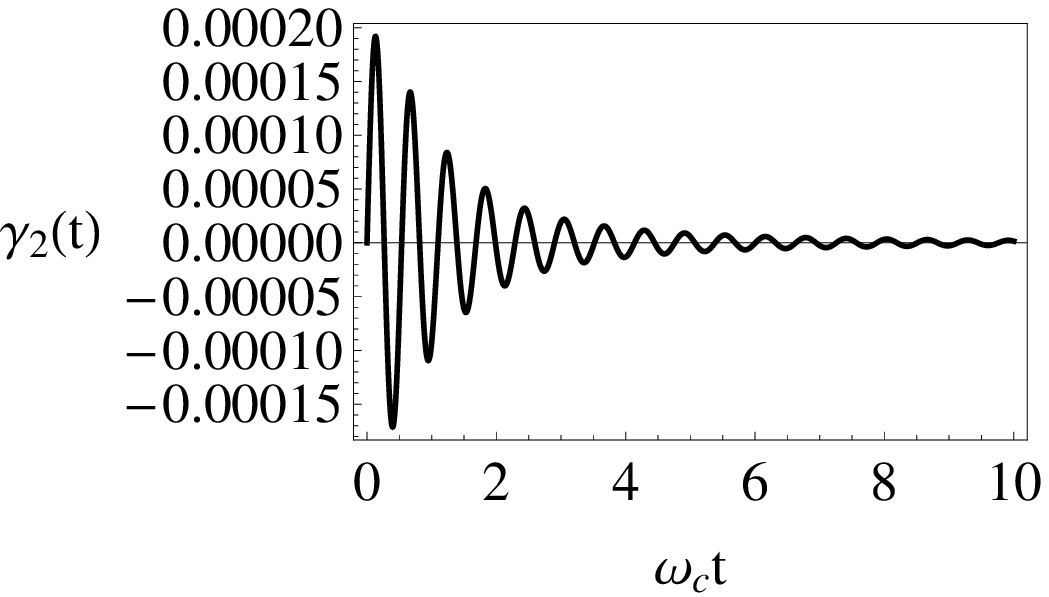}
  \caption{The decay rate $\gamma_2(t)$ in units of $\omega_c$ for $\alpha=0.01$, $\omega_0/\omega_c=10$ and $\epsilon/\Delta=1/(2\sqrt{3})$.}
  \label{figure2}
\end{figure}

\begin{figure}[!h]
    \includegraphics[width=0.4\textwidth]{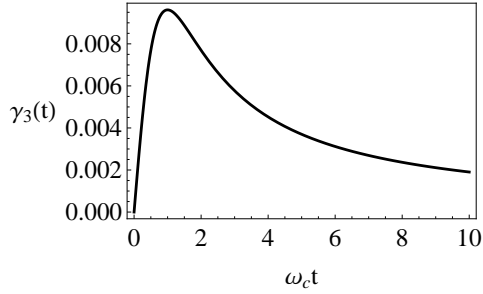}
  \caption{The decay rate $\gamma_3(t)$ in units of $\omega_c$ for $\alpha=0.01$, $\omega_0/\omega_c=10$ and $\epsilon/\Delta=1/(2\sqrt{3})$.}
  \label{figure3}
\end{figure}

The spin-boson model involves three jump operators $C_1=\sigma_-$, $C_2=\sigma_+$ and $C_3=\sigma_z$. The operators $C_1$ and $C_2$ generate jumps that affect the populations of the system eigenstates and the operator $C_3$ produces phase flips. The corresponding decay rates are $\gamma_1(t)=\frac{\Delta^2}{4 \omega_0^2} \gamma_{\omega_0}(t)$, $\gamma_2(t)=\frac{\Delta^2}{4 \omega_0^2}\gamma_{-\omega_0}(t)$ and $\gamma_3(t)=\frac{\epsilon^2}{4 \omega_0^2} \gamma_0(t)$ (c.f. \eref{equation7} and \eref{equation9}), which are plotted in figures \ref{figure1},\ref{figure2} and \ref{figure3}. The first two decay rates have temporarily negative values while the third decay rate remains always positive. This indicates that in channels $C_1=\sigma_-$ and $C_2=\sigma_+$ there occurs reversed jumps resulting into memory effects, but that the phase flip channel $C_3=\sigma_z$ lacks memory.


\begin{figure}[!h]
    \includegraphics[width=0.4\textwidth]{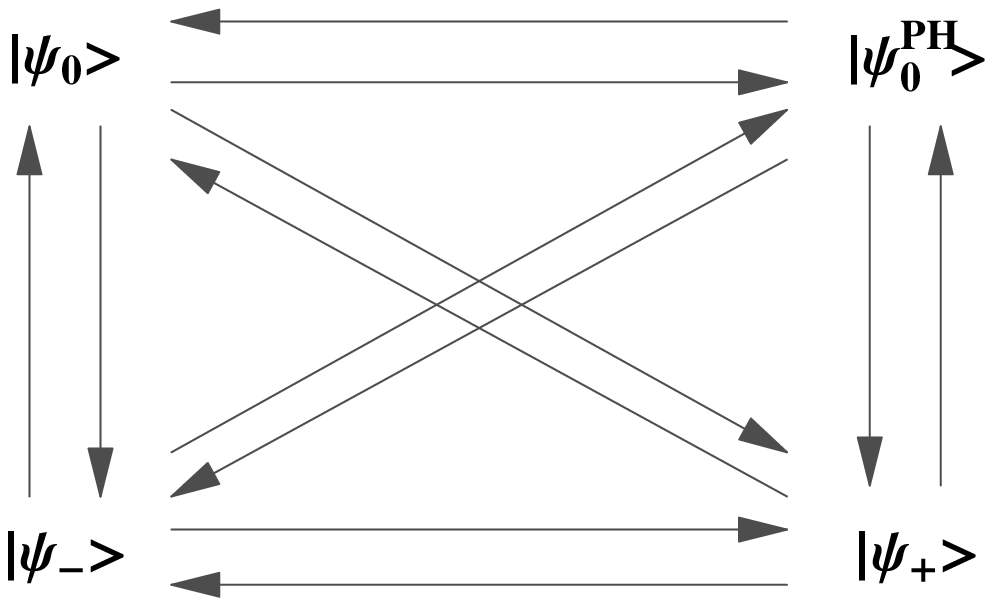}
  \caption{The states involved in the effective ensemble \eref{equation13}. The possible jumps in the dynamics are presented as arrows between the states. Here, $\ket{\psi_0}=\alpha_0\ket{\psi_+}+\beta_0\ket{\psi_-}$ and $\ket{\psi_0^\mathrm{PH}}=\alpha_0\ket{\psi_+}-\beta_0\ket{\psi_-}$.}
  \label{figure4}
\end{figure}

Assume now that the system is initially in a pure state $\ket{\psi_0(0)}=\frac{1}{\sqrt{2}}(\ket{\psi_+}+\ket{\psi_-})$, i.e., all the ensemble members are initially in the same state. We use the notation $\ket{\psi_0}$ for the state vector exposed to only deterministic evolution. When $\epsilon\neq 0$, there exists a phase flipped counter part $\ket{\psi_0^{\mathrm{PH}}}$ for the state $\ket{\psi_0}$, into which the single realization can jump through channel $C_3=\sigma_z$. The possible states and jumps are demonstrated in figure \ref{figure4}.

\begin{figure}[!h]
  {\includegraphics[width=0.4\textwidth]{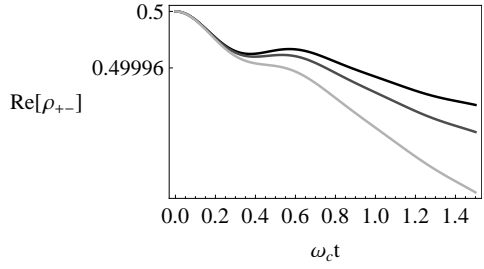}}                    
  \caption{The coherence for $\frac{\omega_0}{\omega_c}=10$ and $\frac{\epsilon}{\Delta}=\frac{1}{2\sqrt{7}}$ (black line), $\frac{\omega_0}{\omega_c}=10$ and $\frac{\epsilon}{\Delta}=0.255$ (dark gray line), $\frac{\omega_0}{\omega_c}=10$ and $\frac{\epsilon}{\Delta}=\frac{1}{2\sqrt{3}}$ (light gray line). Here, the coupling constant is $\alpha=0.01$ and the initial state is $\ket{\psi_0(0)}=\frac{1}{\sqrt{2}}(\ket{\psi_+}+\ket{\psi_-})$.}
  \label{figure5}
\end{figure}

In figure \ref{figure5} the coherence is plotted for a fixed ratio $\omega_0/\omega_c$ and varying values of $\epsilon/\Delta$. One can see that as $\epsilon/\Delta$ varies, the behavior of the coherence changes in such way that for small values of $\epsilon/\Delta$ there is recoherence but as the ratio gets larger the system decoheres with no revivals of coherence. Whether the decay rates $\gamma_1(t)$ and $\gamma_2(t)$ get negative values depends only on the ratio $\omega_0/\omega_c$, so for large values of $\epsilon/\Delta$ the system will decohere although there are reversed jumps in the process. 
Let us study the real part of the coherence $\mathrm{Re}\left[\rho_\pm\right]=\frac{1}{N}\sum_i{\alpha_i\beta_i}$, where $\ket{\psi_0^i}=\alpha_i\ket{\psi_+}+\beta_i\ket{\psi_-}$. Jumps $\ket{\psi_0}\to\ket{\psi_+}$,$\ket{\psi_0}\to\ket{\psi_-}$, $\ket{\psi_0^\mathrm{PH}}\to\ket{\psi_+}$, $\ket{\psi_0^\mathrm{PH}}\to\ket{\psi_-}$ and phase flips reduce the number of terms in the sum presenting the coherence and therefore induce decoherence. Reversed jumps $\ket{\psi_0}\leftarrow\ket{\psi_+}$, $\ket{\psi_0}\leftarrow\ket{\psi_-}$ and $\ket{\psi_0^\mathrm{PH}}\leftarrow\ket{\psi_+}$, $\ket{\psi_0^\mathrm{PH}}\leftarrow\ket{\psi_-}$ produce new terms into the sum, but if there is an equal amount of reversed jumps to $\ket{\psi_0}$ and $\ket{\psi_0^\mathrm{PH}}$ the new terms in the sum cancel each other and no recoherence takes place. So the appearance of recoherence does not only depend on the number of reversed jumps, but also on the proportion of the number of jumps into different target states.

\begin{figure}[!h]
{\includegraphics[width=0.4\textwidth]{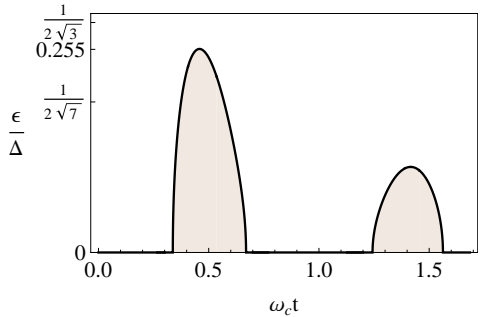}}     
	\caption{The time regions for recoherence for different values of $\frac{\epsilon}{\Delta}$. In light brown areas there is recoherence. Here, $\frac{\omega_0}{\omega_c}=10$, $\alpha=0.01$ and $\ket{\psi_0(0)}=\frac{1}{\sqrt{2}}(\ket{\psi_+}+\ket{\psi_-})$. The values of $\frac{\epsilon}{\Delta}$ used in figure \ref{figure5} are marked in the plot.}
	\label{figure6}
\end{figure}

To find out for which values $\frac{\epsilon}{\Delta}$ there is recoherence let us study the quantity $N_0(t)-N_0^\mathrm{PH}(t)$. Here, $N_0(t)$ is the number of ensemble members in state $\ket{\psi_0}$ and $N_0^\mathrm{PH}(t)$ the number of elements in state $\ket{\psi_0^\mathrm{PH}}$. The real part of the coherence is an increasing function, when $N_0(t)-N_0^\mathrm{PH}(t)$ is increasing, and a decreasing function when $N_0(t)-N_0^\mathrm{PH}(t)$ is decreasing. The regions, where there is recoherence for a fixed value of $\omega_0/\omega_c$ can be seen in figure \ref{figure6}. One can see that as the energy bias $\epsilon$ becomes larger compared with the tunneling amplitude $\Delta$ the probability of a phase flip increases. As more phase flips occur a reversed jump to $\ket{\psi_0^\mathrm{PH}}$ gets as probable as a reversed jump to state $\ket{\psi_0}$, which then causes the system to decohere.  

The study above illustrates how for the spin-boson model there are cases where there are no revivals of coherence, but there occurs reversed jumps indicating memory effects on the level of single wave function realizations. There are no revivals of coherences due to the co-operative action of the phase flip channel and the energy exchanging channels. I.e, although there is information about the phase returning to the system via reversed jumps in the energy exchanging channels, the overall information flow about the phase is from the system to the environment due to the phase flips occurring in the dynamics. This example illustrates, how in complex quantum dynamics it requires subtlety to determine whether memory effects play a role in the system dynamics.

The NMQJ-approach gives a scheme to detect whether there are memory effects in the system dynamics although there are no revivals of coherence in the level of the density matrix. In \cite{nm} the concept of non-Markovianity was clarified by defining a measure for non-Markovianity. It measures the amount of information flowing from the environment to the system. For the spin-boson model this measure is independent of the parameter $\epsilon$, i.e., the maximum value for non-Markovianity is obtained when there are no phase flips occurring in the dynamics.  The NMQJ-description thus gives a similar picture of the memory effects in the dynamics of the spin-boson model as the measure of non-Markovianity.

\section{Conclusions}\label{conclusions}
We derived a time-convolutionless master equation describing the spin-boson model in the weak coupling limit. The equation describes the short time memory effects in the dynamics manifesting themselves through negative decay rates in the master equation.

The memory effects are studied via the non-Markovian quantum jump approach. This description shows memory effects on the level of singe wave function realizations although there are no revivals of coherence in the density matrix description. This behavior demonstrates, how in complex quantum dynamics the co-operative action of different channels complicates the detection of memory effects in the dynamics.

\section*{Acknowledgments}
The author is grateful to Jyrki Piilo for helpful discussions and comments.

\end{document}